\documentclass{IEEEcsmag}

\usepackage[colorlinks,urlcolor=blue,linkcolor=blue,citecolor=blue]{hyperref}

\usepackage{upmath}

\jvol{XX}
\jnum{XX}
\paper{8}
\jmonth{May/June}
\jname{© 2020 IEEE.}
\pubyear{2020}

\begin{document}

%\sptitle{Department: Head}
\editor{Editor: Name, xxxx@email}

\title{Blockchain based Attack Detection on Machine Learning Algorithms for IoT based E-Health Applications}

\author{Thippa Reddy Gadekallu}
\affil{Vellore Institute of Technology, India}

\author{Manoj M K}
\affil{Vellore Institute of Technology, India}

\author{Sivarama Krishnan S}
\affil{Vellore Institute of Technology, India}

\author{Neeraj Kumar*}
\affil{Thapar Institute of Engineering, Punjab \\ School of Computer Science, University of Petroleum and Energy Studies, Dehradun, Uttarakhand }

\author{Saqib Hakak}
\affil{Canadian Institute for Cybersecurity,  Faculty of Computer Science, University of New Brunswick, Canada}

\author{Sweta Bhattacharya}
\affil{Vellore Institute of Technology, India}

%\markboth{Department Head}{Paper title}

\begin{abstract}
The application of machine learning (ML) algorithms are massively scaling-up due to rapid digitization and emergence of new tecnologies like Internet of Things (IoT). In today's digital era, we can find ML algorithms being applied in the areas of healthcare, IoT, engineering, finance and so on. However, all these algorithms need to be trained in order to predict/solve a particular problem. There is high possibility of tampering the training datasets and produce biased results. Hence, in this article, we have proposed blockchain based solution to secure the datasets generated from IoT devices for E-Health applications. The proposed blockchain based solution uses private cloud to tackle the aforementioned issue. For evaluation, we have developed a system that can be used by dataset owners to secure their data.
\end{abstract}

\maketitle

\chapterinitial{introduction} There has been an extensive usage of machine learning (ML) and related applications. This enormous usage has led to reliance on ML based predictions which impacts the decisions taken \cite{AA}. The authenticity of the datasets used to train the machine learning algorithms act as the backbone for such predictions and pertinent decision making. However, if the dataset gets tampered, the training results of the ML algorithms would lead to diluted predictions guiding tilted and biased decisions. As an example, tampering of customer survey research and product review related data could lead to biased product recommendation in e-commerce platforms \cite{AB}. Also, there is a possibility of the ML algorithm being tampered leading to favourable decisions especially in the healthcare sector. 
The capability of ML algorithms to process data and classify data patterns, often increase the susceptibility of ML algorithms towards various types of attacks.  The authors in \cite{AD} have implemented poisoning and evasion attacks with an objective of maximizing the generalization errors in classification. This resulted in a disingenuous model leading to biased values of measurements in classification. Evasion is one of the most commonly practiced attacks where falsified yet normal appearing inputs are fed into the ML algorithm during testing phase. These inputs, when processed, invariably ends up being erroneously classified by the model. The data thus gets misclassified or in some cases leads to concept drift where the system continuously gets retrained, deteriorating the performance \cite{AE}. In case of poisoning attacks, training data gets tampered. This tampered data when fed into a classifier negatively impacts the accuracy of the classification model. In some instances of this type of attack, the classifier function gets skewed producing favourable results for the attacker \cite{AF}, \cite{AG}. With the advancement of several technologies like IoT, Internet of Medical Things, Federated Learning, etc. there is a rapid surge in the digital data in healthcare sector generated through IoT based devices. ML algorithms play a vital role in helping the doctors in diagnosing the patients in a timely manner. Hence, in this article, a blockchain based approach is presented to secure medical datasets generated through IoT devices in healthcare applications.  

The main contribution of this article includes: 

\begin{itemize}
	\item The data generated from the sensors in the IoT system and ML algorithms are stored securely in a private cloud using AES encryption.
	\item Identifying tampering of datasets and ML algorithms using blockchain.
	
\end{itemize}

The organisation of the paper is as follows : Section 2 presents state-of-the-art technologies for securing ML datasets from potential attackers. In section 3, proposed framework is discussed. Section 4 contains experimental results and the article is finally concluded with section 5.

\section{Recent Advances}
The recent studies highlight the exposure of ML algorithms to adversarial attacks where non traceable changes are introduced in the input data leading to erroneous predictions of outputs deceiving the ML algorithm used. The authors in \cite{AH} define and analyse the various forms of adversarial attacks launched in real time situations and also propose the plausible defence strategies to combat such attacks. In case of adversarial images, adversarial noise gets introduced which are used to train machine learning models being subjected to black box attacks. The detectors help to identify adversarial changes incorporated in the original image. The threats relevant to adversarial attacks predominantly exist in classification of image objects captured through cell phone camera where even the google inception model falls prey to such attacks. The Robust Physical Perturbation algorithm is a case wherein imposters print forged road sign posters and replace it with the real sign. Similar discrepant approaches have been identified in the form of cyberspace attacks. Robotic visual images and also three dimensional object images are fed to ML algorithms for classifications and predictions \cite{AI}, \cite{AJ}. One of the most interesting applications of blockchain  is intrusion detection. Intrusion detection with  the intersection of blockchain  has  huge  scope of implementation in case of cryptocurrency and smart contract \cite{AK}. Blockchain has a lot of potential applications in the energy sector which can be observed in peer to peer energy trading, IoT applications incorporating blockchain, decentralized marketplaces, charging of electric vehicles and e-mobility \cite{AL}. The non- financial applications of blockchain are Ethereum and Hyperledger. The authors in \cite{AP} have identified Binary Neural Network (BNN) to be more robust than full precision networks. Hence input discretization or dimensionality reduction of the input parameters when combined with BNN makes the model more robust against adversarial attacks. 
The challenges of the existing works are summarized in Table \ref{tab:my-table}.
The existing solutions against various types of attacks on training and ML algorithms are given below:

\begin{enumerate}
	\item Adversarial Attack on training data: Adversarial training using Brute Force, Data compression as a counter-measure, Foveated Imaging Mechanism, Randomization of Data
	\item Adversarial Attack for network model: Deep Contractive Network, Regularization and Masking of the Gradient, Defensive Filtration, Bioinspired Defence Mechanism
	\item Poisoning Attack: Sanitization of Data, Micromodel based defence, Strong Intentional Perturbations,Human in the Loop (HITL) model, TRIM algorithm
	
\end{enumerate}

\begin{table*}[]
	\centering
	\caption{Summary of the challenges in existing literature.}
	\label{tab:my-table}
	\resizebox{\textwidth}{!}{%
		\begin{tabular}{|l|p{4.5cm}|p{3.5cm}|p{5cm}|}
			\hline
			\multicolumn{1}{|c|}{\textbf{Ref.}} &
			\multicolumn{1}{c|}{\textbf{Methods used}} &
			\multicolumn{1}{c|}{\textbf{Evaluation metrics}} &
			\multicolumn{1}{c|}{\textbf{Research Challenges}} \\ \hline
			\cite{AB} &
			Blockchain technology to secure e-commerce transactions &
			MD5, smart contracts and digital signatures &
			Scalability, computing resources \\ \hline
			\cite{AF} & Linear regression                        & Mean Squared Error, Execution time & Delay/overhead in data processing        \\ \hline
			\cite{AK} & Intrusion Detection System on Blockchain & Data integrity, transparency       & Attacks prevention, scalability          \\ \hline
			\cite{AP} & Binary neural networks                   & Weight decay value, learning rate  & Multi-steps attacks still occur          \\ \hline
			\cite{AR} & Blockchain system for dApps              & Smart contracts                    & Transaction delay, lacks high throughput \\ \hline
		\end{tabular}%
	}
\end{table*}

Some of the limitations of the existing defence mechanisms include:
\begin{enumerate}
	\item The existing defence mechanisms deal with specific type of attacks and hence fail to adapt to newer attacks.
	\item The defence mechanisms such as Brute force method consumes excessive  computational resources.
\end{enumerate} 

The present work emphasizes on elimination of these limitations using the proposed blockchain based approach.

\section{Blockchain-Based Fragmentation Approach to Secure Machine Learning Datasets}

\subsection{Background}
Blockchain is a technology wherein list of timestamped immutable data records are stored and managed in blocks by groups of computational entities. The blocks are interconnected with one another through cryptographic hashes of the preceding block. Each block contains three components namely the timestamp, hash of the preceding block and the data pertaining to transactions. Hence if any updates in the transactions need to incorporated, it has to be uniformly updated in all the blocks constituting the blockchain through consensus mechanism \cite{AQ}. This ensures immutability property of the blockchain which establishes blockchain as the ideal technology for addressing all types of attacks on ML algorithms. Blockchain technologies have been successfully implemented in cryptocurrencies, supply chain management, asset management, health care, maintenance of digital Ids and many others \cite{AR,15}. With the present world being dependent on data centric analysis requiring accurate ML algorithms, it becomes extremely necessary to ensure defence from all possible attacks. There is dire need to build a model that is robust enough to combat all such attacks on datasets and ML algorithms. This acts as the primary motivation behind the present work conducted. In this work, the datasets and the ML algorithm are stored in an encrypted format in the private cloud. Any user who intends to use this dataset or ML algorithm will be issued a block id along with the hash of the dataset. The user on receipt of this id can apply the ML algorithms on the datasets to perform predictive analysis. On completion of this process at the user end, a new hash will be generated. This user generated hash will be compared with the hash of the blockchain. If the hashes match it can concluded that the datasets and the ML algorithms have not been compromised.

\subsection{Proposed Architecture}
\begin{figure}[h!]\centering
	\frame{\includegraphics[width= 7cm]{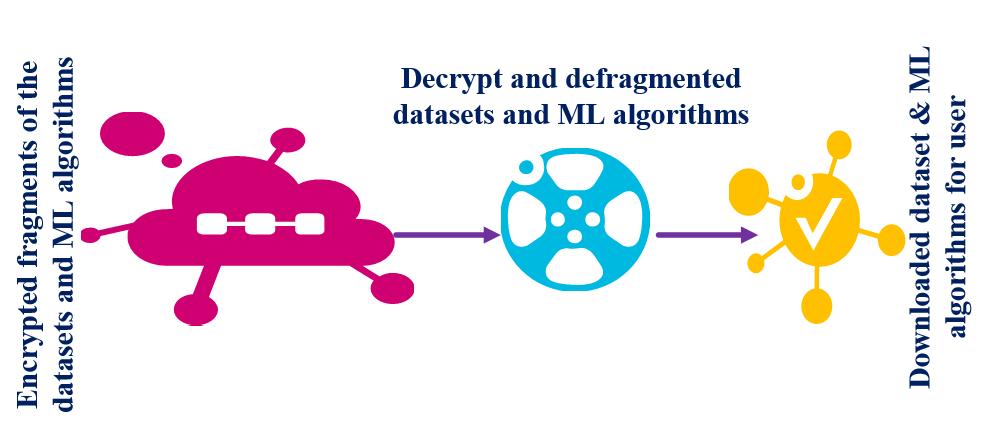}}
	\caption{Storage of Fragmented dataset in Private Cloud}
	\label{fig:Fig 1}
\end{figure}

\begin{figure} [h!]\centering
	\frame{ \includegraphics[width=7cm]{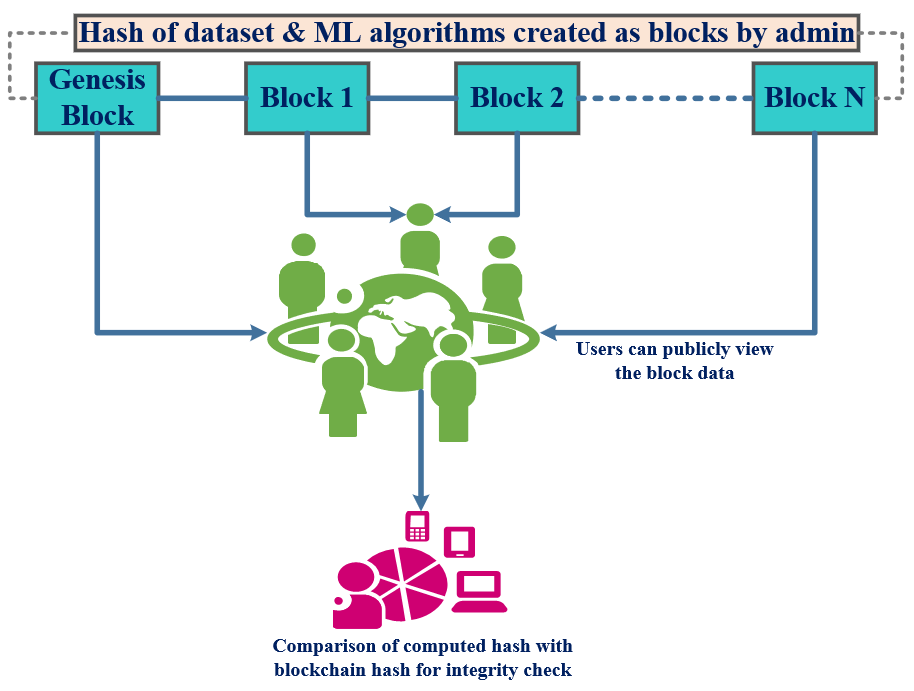}}
	\caption{Storage of encrypted fragments in Blockchain and verification of hash after experimentation}
	\label{fig:Fig 2}
\end{figure}

Figures \ref{fig:Fig 1} and \ref{fig:Fig 2} represent the architecture diagram of the proposed work. 
Figure \ref{fig:Fig 1} describes the process of data handling and storage in private cloud. The private cloud holds the encrypted fragments of the data sets and ML algorithms. When user initiates a request for download, the data is decrypted and defragmented.
Figure \ref{fig:Fig 2} describes the hybrid blockchain wherein the creation of the blocks is done by the administrator depicting the private blockchain and the visibility and access of the blocks are provided to the user representing the public blockchain. 


The key objective in using a private cloud is to let the owner have full control over the dataset. The owner uses the private cloud to restrict access to approved users and eliminate unauthorized access. This level of access control greatly improves the overall objective of securing the dataset. On the cloud, the dataset is stored as encrypted fragments so as to improve security. Once a download request for the dataset is initiated, the fragments are decrypted and defragmented so as to provide the user the original dataset file. 
\\

The user may then use the public blockchain to view and verify the hash of file with the computed hash of the downloaded file to ensure file integrity. This helps to establish and justify the integrity of the dataset to any third party.
\\

The admin is responsible to add the dataset name and hash of the file into the blockchain. This is done with the special private blockchain access through the admin private key wherein he/she may add a dataset hash as a block to the blockchain making it visible publicly thereby maintaining integrity of the file.
\\

This form of integrity check with a blockchain brings in a new flavour to the existing forms of security and can act as a stepping stone for more futuristic ideas of automated security. The hybrid blockchain can act as a means of utilizing the features of both private and public blockchain to get a desired outcome. Here, we bring in the concept of full authority to the owner of the data while not restricting view of the data to the public.

\section{Experiments and Results}

\begin{figure}[h!]\centering
	\frame{ \includegraphics[width=7cm]{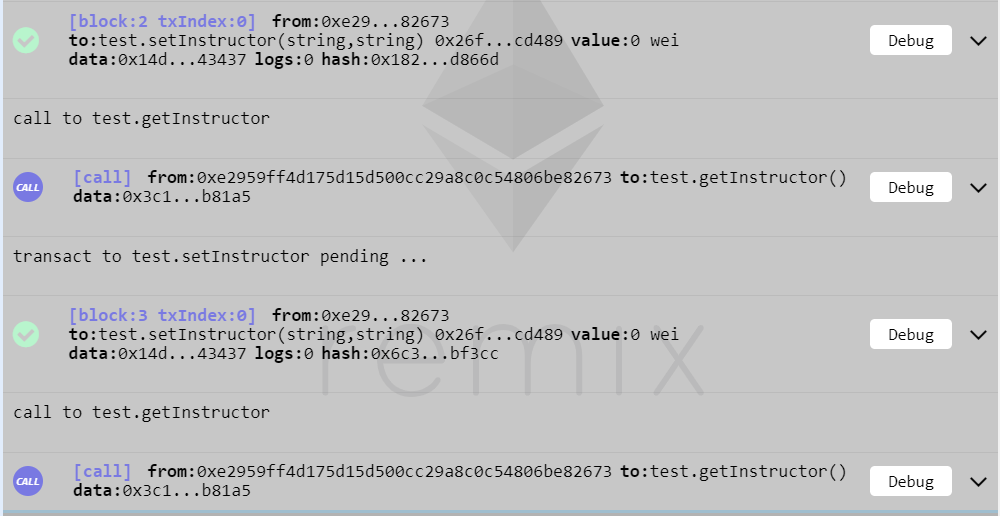} } 
	\caption{Log of the Blocks created in Blockchain}
	\label{fig:Fig 8}
\end{figure}

To simulate the experimentation, the following software are used in this work. For fragmentation we have used 7Zip, an open source file archiever software. The private cloud is hosted in Google Cloud Platform. Blockchain is simulated with the help of Remix IDE (Ethereum) through smart contract developed using Solidity. To conduct this experimentation, Medical Cost Dataset from Kaggle is used. This dataset has 1338 Rows of data with 7 Attributes. Before storing the dataset in private cloud, it has been divided into several fragments using 7zip open source file archiever software. These fragments are then encrypted using AES encryption with 256 Bit Key size and uploaded to the Virtual Private Cloud (VPC) in Google Cloud. The admin can then compute hash of the datasets and ML algorithm and store the same in a blockchain.  The linear regression algorithm is used for experimentation purposes in the present study.  The sample logs created in the block chain is depicted in Figure \ref{fig:Fig 8}.

 A simulation of the deployed contracts is performed to manage the blocks in the blockchain.
 
 If the user wants to test the accuracy of ML algorithm on the dataset, he can request access from the admin for the same. When the user provides a private key, the dataset will be defragmented and the user can download the dataset and ML algorithm. The user may compute the hash of the file downloaded and compare the hash with the public blockchain access following which the experimentation of ML algorithms
 on the dataset can be performed by the user. After experimentation, any third party may verify the originality of the results obtained by comparing the generated hash with the public blockchain hash. If the hashes match, it means that the dataset and ML algorithm is not compromised.

\section{Conclusion and Future Scope}
In this work we successfully implemented a blockchain based solution to identify attacks on ML algorithms and medical datasets. The use of the same concept to power the need of securing datasets of an organization would mean that the private blockchain requires authentication from a wide range of higher officials awaiting a consensus.  A feasibility check on the different consensus for such a large scenario while taking into consideration the processing power, time and resources for data block creation and mining could be a much needed analysis. 
A complete decentralized solution of this could be the use of decentralized storage such as Inter Planetary File System (IPFS) or SWARM so that the dataset may be kept more secure and not on a single entity. Securing the dataset by decentralised storage may be a stepping stone in to the future of decentralization, a peek into web 3.0.

\begin{IEEEbiography} {Thippa Reddy Gadekallu}{\,}is currently working as an Associate Professor in School of Information Technology and Engineering, Vellore Institute of Technology, Vellore, Tamil Nadu, India. He obtained his Bachelor of Technology degree in Computer Science and Engineering from Nagarjuna University, Andhra Pradesh, India, Master of Engineeting in Computer Science and Engineering from Anna University, Chennai, Tamil Nadu, India and completed his Ph.D in Vellore Institute of Technology, Vellore, Tamil Nadu, India. He has 14 years of experience in teaching. He coauthored more than 80 international publications. Currently, his research interests include Machine Learning, Deep Learning, Computer Vision, Big Data Analytics, Blockchain.  
\end{IEEEbiography}

\begin{IEEEbiography}{Manoj MK}{\,} is currently working in Oracle India Pvt. Ltd., India. He has completed his  Master of Technology in Software Engineering in Vellore Institute of Technology, India.  He has done various projects on Block Chain, Cloud Security, Machine Learning, AI and IOT. He has been awarded FTRI (Fast Track Research Initiative) -  G D Naidu young Scientist Award from VIT. He has published a chapter for a book on blockchain. His interest lies deeply on futuristic technologies. Contact him at mkmanoj1997@gmail.com.
	
\end{IEEEbiography}

\begin{IEEEbiography}{Siva Rama Krishnan S}{\,} is currently working as an  Assistant Professor in Vellore Institute of Technology, India. He was a research member in Centre for Ambient Intelligence and Advanced Networking Research (AMIR). He has working experience in Centre for Development and Advance Computing (C-DAC) (Ministry of Science and Technology, Govt. of India), as research intern in data center technologies. He is also certified by EMC corp. as a proven professional in Information Storage and Management. Currently an EMC academic alliance faculty and played a key role in establishing a MoU between VIT University and EMC corp. He had proposed and developed an Intelligent Network design framework for building small and large-scale network. He also developed an efficient and secure framework for IP Storage network for C-DAC.  His current interests includes e-waste management in India, wireless networks and cloud computing.

\end{IEEEbiography}

\begin{IEEEbiography}{Neeraj Kumar}{\,} received his Ph.D. in CSE from Shri Mata Vaishno Devi University, Katra (Jammu and Kashmir), India in 2009, and was a postdoctoral research fellow in Coventry University, Coventry, UK. He is working as a Professor in the Department of Computer Science and Engineering, Thapar Institute of Engineering and Technology (Deemed to be University), Patiala (Pb.), India. He is also with School of Computer Science, University of Petroleum and Energy Studies, Dehradun, Uttarakhand. He has published more than 500 technical research papers in top-cited journals such as IEEE TKDE, IEEE TIE, IEEE TDSC, IEEE TITS, IEEE TCE, IEEE TII, IEEE TVT, IEEE ITS, IEEE SG, IEEE Netw., IEEE Comm., IEEE WC, IEEE IoTJ, IEEE SJ, Computer Networks, Information sciences, FGCS, JNCA, JPDC and ComCom. He has guided many research scholars leading to Ph.D. and M.E./M.Tech. His research is supported by funding from UGC, DST, CSIR, and TCS. He is an Associate Technical Editor of IEEE Communication Magazine. He is an Associate Editor of IJCS, Wiley, JNCA, Elsevier, Elsevier Computer Communications, and Security and Communication, Wiley. He has been a guest editor of various International Journals of repute such as - IEEE Access, IEEE Communication Magazine, IEEE Network Magazine, Computer Networks, Elsevier, Future Generation Computer Systems, Elsevier, Journal of Medical Systems. Springer, Computer and Electrical Engineering, Elsevier, Mobile Information Systems, International Journal of Ad hoc and Ubiquitous Computing, Telecommunication Systems, Springer and Journal of Supercomputing, Springer. He has been a workshop chair at IEEE Globecom 2018 and IEEE ICC 2019 and TPC Chair and member for various International conferences. He is senior member of the IEEE. He has more than 20,000 citations to his credit with current h-index of 77. He has won the best papers award from IEEE Systems Journal and ICC 2018, Kansas city in 2018. He is visiting research fellow at Coventry University, Newcastle University, UK.

\end{IEEEbiography}

\begin{IEEEbiography}{Saqib Hakak}{\,} is currently working as an Assistant Professor, Canadian Institute for Cybersecurity,  Faculty of Computer Science,  University of New Brunswick,  Fredericton,  Canada. He received his Ph.D. from the University of Malaya, Malaysia, under the Faculty of computer Science and Information Technology. He received his Bachelor’s degree in computer science engineering from the University of Kashmir, India, in 2010 and his Master’s degree in Computer and information engineering from IIUM, Malaysia. His research areas include information security, natural language processing, cyber security, artificial intelligence, and wireless networks.
	
\end{IEEEbiography}

\begin{IEEEbiography}{Sweta Bhattacharya}{\,} is currently associated with Vellore Institute of Technology (University), as an Assistant Professor in the School of Information Technology \& Engineering. She has received her PhD degree from Vellore Institute of Technology and Master’s degree in Industrial and Systems Engineering from State University of New York, Binghamton, USA. She has guided various UG and PG projects and published peer reviewed research articles. She is also a member of the Computer Society of India and Indian Science Congress. Her research experience includes working on Pill Dispensing Robotic Projects, as a fully funded Watson Research Scholar at Innovation Associates, Binghamton at SUNY. She has completed six sigma green belt certification from Dartmouth College, Hanover. Her research interests include applications of machine learning algorithm, data mining, simulation and modelling, applied statistics, quality assurance and project management. Contact her at sweta.b@vit.ac.in.
	
\end{IEEEbiography}

\end{document}